 \pdfoutput=1
\documentclass[%
preprint,
 amsmath,amssymb,
 aps,
floatfix,
]{revtex4-1}
\pdfoutput=1
\usepackage{appendix}
\usepackage{amsmath}
\usepackage{graphicx}
\usepackage{epstopdf}
\usepackage{graphics}
\usepackage{epsfig}
\usepackage[byname]{smartref}
\usepackage{hyperref} 
\usepackage{txfonts}
\usepackage{morefloats}
\usepackage{tocloft}
\usepackage{graphicx}
\usepackage{dcolumn}
\usepackage{bm}
\usepackage{hyperref}

\begin{document}

\preprint{V1}

\title{Understanding of hopping matrix for 2D materials taking 2D honeycomb and square lattices as study cases}

\author{Maher Ahmed}
\affiliation{Department of Physics and Astronomy, University of Western Ontario, London ON N6A 3K7, Canada}
\affiliation{Physics Department, Faculty of Science, Ain Shams University, Abbsai, Cairo, Egypt}
\email{mahmed62@uwo.ca}
%



\begin{abstract}
In this work, a trial understanding for the physics underling the
construction of exchange (hopping) matrix $\mathbf{E}$ in Heisenberg model
(tight binding model) for 2D materials is done. It is found that the
$\mathbf{E}$ matrix describes the particles exchange flow under short range
(nearest neighbor) hopping interaction which is effected by the lattice
geometry. This understanding is then used to explain the dispersion relations
for the 2D honeycomb lattice with zigzag and armchair edges obtained for
graphene nanoribbons and magnetic stripes. It is found that the particle flow
by hopping in the zigzag nanoribbons is a translation flow and shows
$\mathbf{\cos^2}(q_xa)$ dependance while it is a rotational flow in the
armchair nanoribbons. At $q_xa/\pi=0.5$, the particles flow in the edge sites
of zigzag nanoribbons with dependance of $\mathbf{\cos^2}(q_xa)$ is equal to
zero. At the same time there is no vertical hopping in those edge sites which
lead to the appearance of peculiar zigzag flat localized edge states.
\end{abstract}

\pacs{Valid PACS appear here}
\maketitle

\def\baselinestretch{1.66}

\section{Introduction}
\begin{figure}[h]
\centering
\includegraphics[scale=0.2]{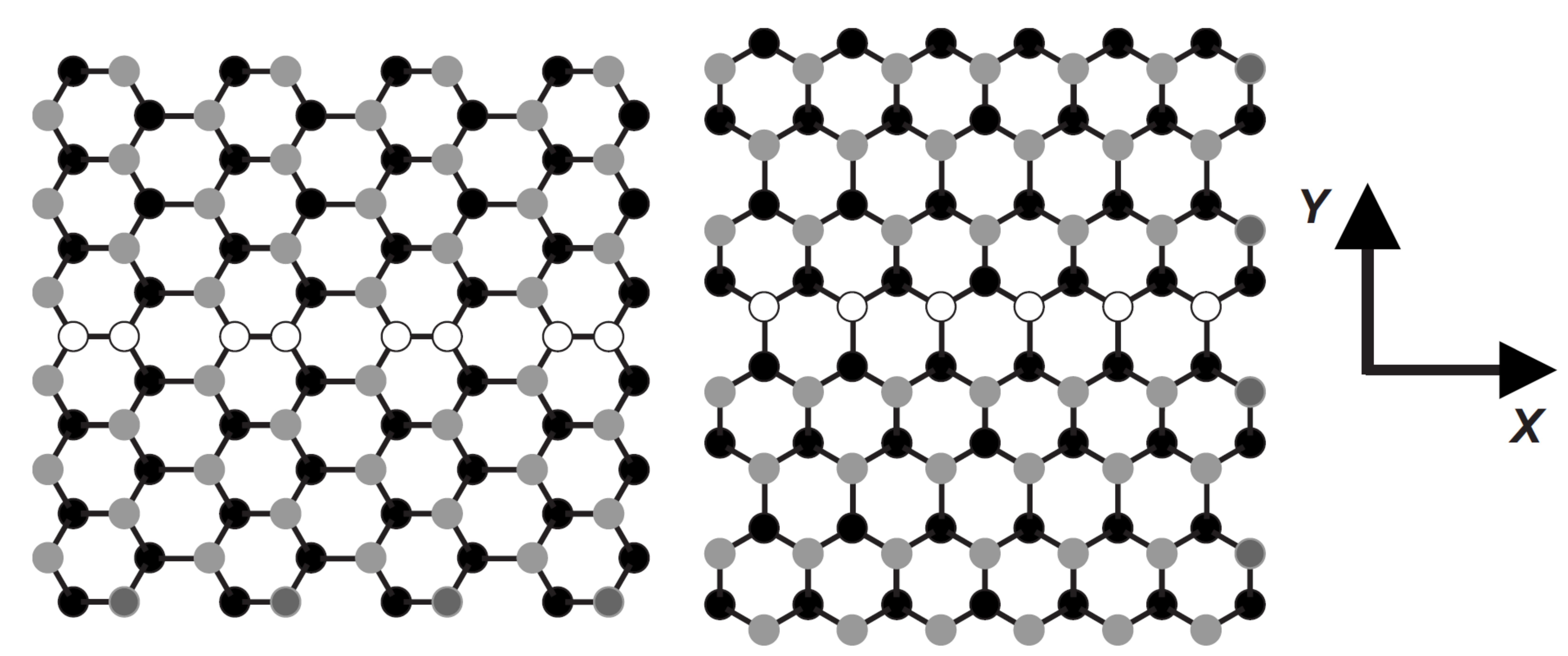}
\caption{Armchair (left) and zigzag (right)  2D Heisenberg ferromagnetic dots honeycomb stripes in $xy$-plane,  where black (gray) dots are the sublattice
A(B) with a line of impurities (white dots) in the middle of
the sheet, and with average spin $S$ alignment in $z$ direction. The stripes are finite in $y$ direction with $N$ rows
($n = 1,\cdots,N$) and they are infinite in the $x$ direction.  Figure taken from \cite{rim1}.}\label{fig:graphenelattice3}
\end{figure}

It is shown in \cite{Selim2011,Ahmed2,rim1} that the physics of the Heisenberg Hamiltonian system and
tight binding Hamiltonian system for 2D honeycomb armchair and zigzag nanoribbons shown in Figure \ref{fig:graphenelattice3} are nearly
equivalent which is a reflection of their equivalent from geometrical and
topological point of view, as both system represent an exchange (a hopping)
flow of particles, electrons (fermions) in graphene case and magnons (bosons)
in magnetic case, under short range interaction (nearest neighbor exchange $J_{ij}$ for magnetic excitations and nearest neighbor hopping $t_{ij}$ for electronic excitations)  through the
same 2D honeycomb lattice.

All the important geometrical and topological
information that effect this exchange (hopping) flow for both systems are
encoded in the following $\mathbf{E}$ matrix \begin{equation}\label{ematrix}
\mathbf{E}=\left[%
\begin{array}{cc}
  \alpha I_N  & T(q_x)  \\
  T^*(q_x) & \alpha I_N  \\
\end{array}%
\right],
\end{equation}
here, $T(q_x)$ is the exchange matrix, which depends on the orientation of the ribbon and is given by
\begin{equation}\label{}
\left(
  \begin{array}{ccccc}
   \varepsilon & \beta   &     0 & 0 & \cdots \\
    \beta   & \varepsilon & \gamma& 0 & \cdots \\
    0       & \gamma  &\varepsilon&  \beta    &\cdots \\
    0       &      0  & \beta & \varepsilon & \cdots \\
     \vdots & \vdots& \vdots & \vdots & \ddots \\
  \end{array}
\right),
\end{equation}
where the parameters $\varepsilon$, $\gamma$, and $\beta$ depend on the stripe edge geometry
and are given in Table \ref{tabexchn}.
\begin{table}[h]
\caption{Nearest neighbor exchange matrix elements for 2D magnetic honeycomb lattice}\label{tabexchn}
\begin{tabular}{lcl}
  \hline\hline
    Parameter     &   Zigzag                  &   Armchair                 \\\hline
   \hspace{50pt}  &  \hspace{50pt}            &  \hspace{50pt}             \\
   $\varepsilon$  &         0                 &$\frac{SJ}{2}e^{-iq_xa}$   \\[10pt]
   $\beta$        &$SJ \cos(\sqrt{3}q_x a/2)$&$\frac{SJ}{2}e^{iq_xa/2}$  \\[10pt]
   $\gamma$       & $\frac{SJ}{2}$           &$ \frac{SJ}{2}e^{iq_xa/2}$  \\
 \hline
\end{tabular}
  \centering
\end{table}

It turns out that allowed
exchange (hopping) flow modes inside the lattice are the eigenvalues of that matrix \cite{Selim2011,Ahmed2,rim1}
\begin{eqnarray}
 \omega(q_x)\left[%
\begin{array}{c}
a_{n}   \\
b_{n}   \\
\end{array}%
\right]  &=& \left[%
\begin{array}{cc}
  \alpha I_N  & T(q_x)  \\
  T^*(q_x) & \alpha I_N  \\
\end{array}%
\right] \left[%
\begin{array}{c}
a_{n}   \\
b_{n}   \\
\end{array}%
\right], \label{eq3eqing}
\end{eqnarray}
where $a_{n}$ and $b_{n}$ are the annihilation boson
operators for the 2D honeycomb sublattices A and B respectively \cite{Neto1}, $q_x$  is wavevector along the $x$ axis which is the translation
symmetry direction of the nanoribbons, and $\omega(q_x)$ are the frequencies of the spin wave modes.

The $\mathbf{E}$ matrix describes in general two allowed directions for particles
exchange (hopping) flow: one along the direction of translation symmetry for
the 2D lattices, and the other along the vertical to that translation
symmetry direction.  The main effect of particles exchange (hopping) flow
along the direction of translation symmetry for the 2D lattices is the
changing in the energy of allowed propagation modes due to the 2D lattice
symmetry encoded as a function in the particles momentum component along that
direction of translation symmetry. The main effect of vertical particles
exchange (hopping) flow in the 2D lattice stripes and nanoribbons is the
quantization of allowed modes due to the quantum confinement effect for
particles motion in the vertical direction to translation symmetry axis. This
vertical particles exchange (hopping) flow is independent of the particles
momentum component in the direction of translation symmetry for the 2D
lattices.

\section{Understanding exchange matrix}
The $\mathbf{E}$ matrix has two sub matrixes components: $\alpha I_N$ and $
T(q_x)$. The first sub matrix component $\alpha I_N$ represents insite energy
value in the lattice, which in turn represent each sites potential energy for
exchange flow of particles inside the lattice. When all sites have the same
potential energy  value, (i.e. perfect and impurity free lattice), the
resistance for exchange flow between the lattice sites is nearly zero and
consequently the particles flow form a perfect fluid, which can be seen in
graphene \cite{PhysRevLett.103.025301}. Introducing any change for insite
energy in the lattice for example the effects due to change edge uniaxial
anisotropy studied in \cite{Ahmed2} resulting changing in edges insite energy
which break the flow symmetry in the lattice as it is seen in magnetic
stripes and graphene nanoribbon \cite{PhysRevB.73.045432}.

The second sub matrix component $T(q_x)$ represents the effect of lattice
geometry in the particles exchange flow (propagation) inside the lattice
under nearest neighbor exchange (hopping) which depends on the edge
configuration as zigzag or armchair \cite{PhysRevB.59.8271}. To further
clarify the above meaning of the $T(q_x)$ matrix, a closer examination of its
derivation in \cite{Selim2011,Ahmed2,rim1} is needed. Its derivation starts from the following exchange sum
\begin{equation}\label{exsum4}
\gamma(q_x) = \frac{1}{2} S\sum_{\nu} J_{i,j}  e^{-i\mathbf{q}_x \cdot (\mathbf{r}_i-\mathbf{r}_j)}.
\end{equation}
The sum for the exchange terms $J_{i,j}$ is taken to be over all $\nu$ nearest neighbors
in the lattice which depends on the edge configuration as zigzag or armchair for the stripe.
For the armchair configuration, the exchange sum gives the following amplitude factors $ \gamma_{nn'}(q_x)$
\begin{eqnarray}\label{amchairaf}
   \gamma_{nn'}(q_x)=\frac{1}{2} SJ\left[\exp(iq_xa)\delta_{n',n}+\exp\left(i\frac{1}{2}q_xa\right)\delta_{n',n\pm1}\right],
\end{eqnarray}
while for the zigzag case it gives
\begin{eqnarray}\label{zigzagaf}
  \gamma_{nn'}(q_x)=\frac{1}{2} SJ\left[2\cos\left(\frac{\sqrt{3}}{2}q_xa\right)\delta_{n',n\pm1}+\delta_{n',n\mp1}\right].
\end{eqnarray}
The $\pm$ sign depends on the sublattice
since the sites line alternates from A and B.

The exchange sum represents the directed component of exchange flow to
each nearest neighbor with respect to the direction of translation symmetry
of the stripe described by Fourier transform. Applying the exchange sum
\ref{exsum4} to each armchair and zigzag site with its nearest neighbor
connections in 2D honeycomb lattice in the direction of translation symmetry
of the stripe as shown in Figure \ref{sites} results in the amplitude factors
\ref{amchairaf} and \ref{zigzagaf}, which are the elements of exchange matrix
$T(q_x)$. Each element in this matrix is the product of exchange strength and
geometrical amplitude as seen in Table \ref{tabexchn}, which expressing the
modulation of nearest neighbor exchange strength due to the flow topology
inside the lattice which depend on both the wavevector (i.e the momentum) of
the particle and the edge configuration as zigzag or armchair.

The matrix elements consistent of three types: the diagonal element
representing the nearest neighbor exchange between sites lies in the same
line along the direction of translation symmetry of the stripe, after
diagonal element representing the nearest neighbor exchange between the sites
at the same line and next line in the lattice sites, and before diagonal
element representing the nearest neighbor exchange between the sites in same
line and upper line in the lattice sites.

\section{Applying exchange matrix to 2D Honeycomb Lattice}
\begin{figure}[h]
\centering
  \begin{tabular}{cc}
\includegraphics[scale=.6]{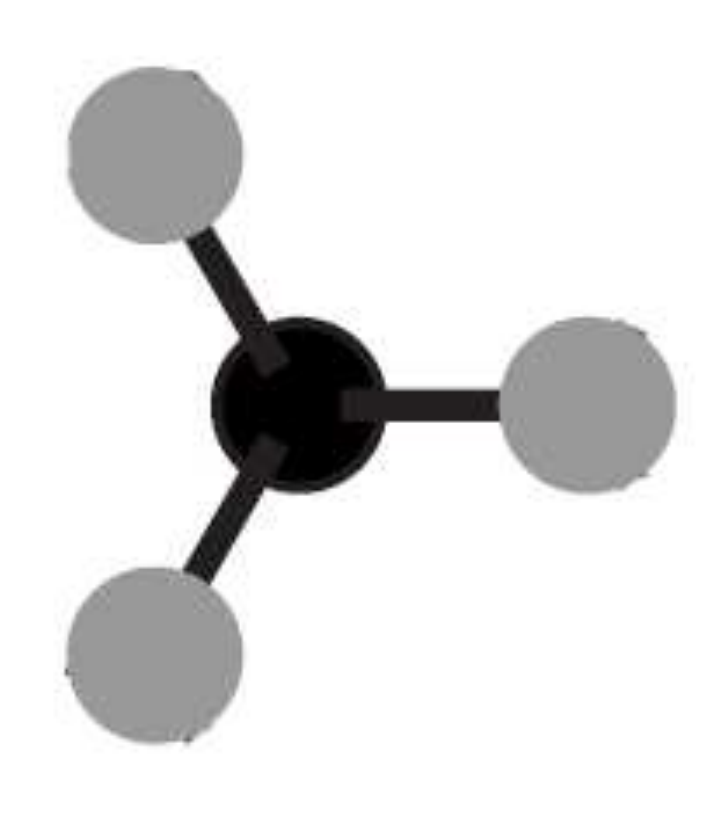}& \includegraphics[scale=.6]{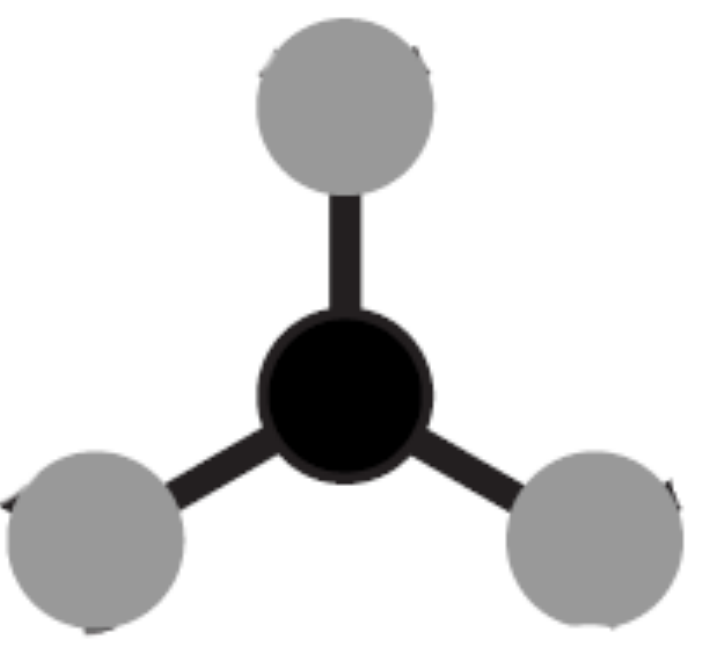}\\
\includegraphics[scale=.5]{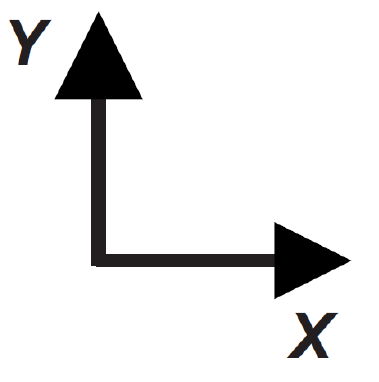}& \\
\end{tabular}
  \caption{Nearest neighbor connections for a site in 2D honeycomb lattice in the direction of translation symmetry of the stripe. The Right is the armchair site while the left is the zigzag site}\label{sites}
\end{figure}

Using the above explanation for the elements of exchange matrix $T(q_x)$ and
the Table \ref{tabexchn}, we can now understand the exchange (the hopping)
flow of particles in 2D Honeycomb Lattice.  Beginning by the zigzag stripes,
the diagonal elements is zero since the sites in the same line in  zigzag
stripe are not nearest neighbor and therefore no exchange flow through that
line. For up and under diagonal elements, the alternates between A and B
sites lines create alternates parallel connected zigzag lines with vertical
connections, which clear from Figures \ref{fig:graphenelattice3} and
\ref{sites}.

The element $\beta$ represents the exchange flow in the parallel zigzag lines
along the translation symmetry of the zigzag stripe, where the real term
$[2\cos(\sqrt{3}q_x a/2)]$ comes from the sum of $[\exp(-i\sqrt{3}q_x
a/2)+\exp(i\sqrt{3}q_x a/2)]$ which reflect the ability to move nearly linear
parallel to the translation symmetry direction, which modulate the exchange
strength according to the particle momentum $q_x$. The element
$\gamma=({SJ}/{2})$ represents the exchange flow in vertical connections
between the parallel zigzag lines and perpendicular to the translation
symmetry of the zigzag stripe, the term comes form
$[({SJ}/{2})\exp(-i\mathbf{q}_x \cdot (\mathbf{r}_i-\mathbf{r}_j))]$, which
is equal to $({SJ}/{2})$  since $(\mathbf{r}_i-\mathbf{r}_j)$ is
perpendicular to $q_x$ for vertical sites which leads the exponential term to
be equal to 1, and therefore the exchange strength in the vertical direction
is constant and independent on $q_x$.

\begin{figure}[hp]
\centering
\includegraphics[scale=1]{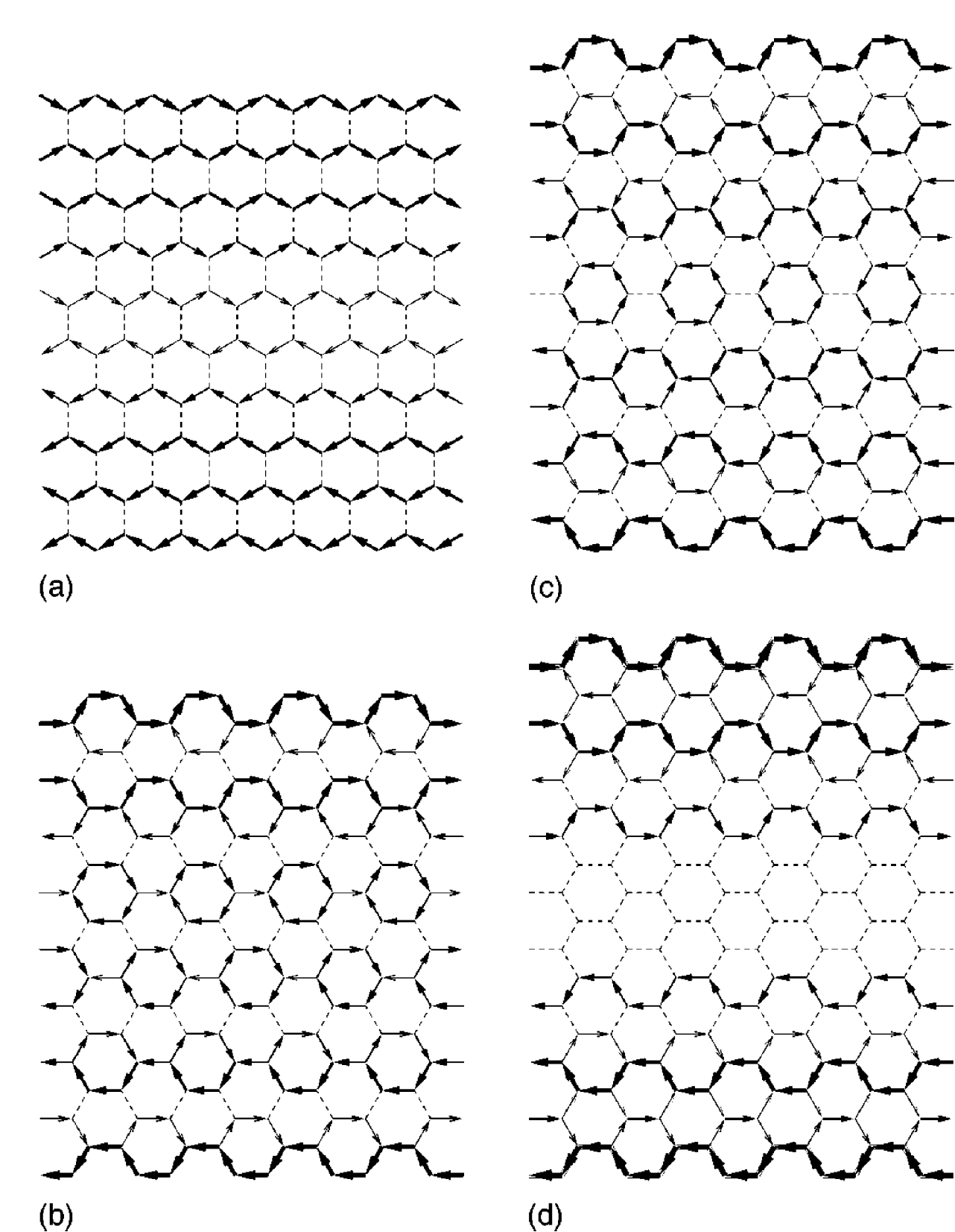}
\caption{The texture of the ring currents under applying week magnetic field perpendicular to graphene nanoribbons for (a) zigzag ribbon
(N=10) and armchair ribbons of (b) N=18, (c) N=19, and (d)
N=20. In zigzag ribbon, because of the symmetry of the lattice, the
ring currents along the vertical bonds are zero. In armchair ribbons
of N=18 and 19, the Kekul\'{e} pattern is clear.  Figure and caption taken from \cite{PhysRevB.59.8271}.} \label{current2}
\end{figure}

The particle in any interior site in the zigzag stripe will be under two
competitive exchange (hopping) force with different strength: one through a
zigzag line along the translation symmetry of the stripe and the other
through vertical connections between the parallel zigzag lines, the main
factor that detriment which direction the particle has high probability to
flow is its momentum in translation symmetry direction $q_x$. The exchange
(hopping) strength in the zigzag lines direction is much larger than the
exchange strength in the noncontinuous vertical lines direction in most of
$q_x$ values and the particle has high probability to flow in zigzag lines.
The direction of flow in upper edge is $x$ direction while in the lower edge
is $-x$ direction (see Figure \ref{current2}a) this is due to the reversing
in the zigzag lines sequence between up and lower edges, i.e.  AB, BA,
AB,....AB, BA. This behavior is displayed in the determinant condition of equation
\ref{eq3eqing} as a dependence on the exchange matrix squared $T^2(q_x)$
\cite{algebra} which leads to $\cos^2(\sqrt{3}q_x a/2)$ dependance of the
modes dispersions of zigzag stripe. When particle momentum $q_x$ is zero
which verify the conditions $q_xa=0$ the exchange (hopping) strength in the
zigzag lines direction is nearly double exchange strength in noncontinuous
vertical direction and the particle has high probability to flow in zigzag
lines which shown as maximum (minimum) energy in the dispersion relations. As
particle momentum reaches the value that verify the condition $q_xa/\pi=0.5$
the exchange (hopping) strength in the zigzag line direction is nearly zero
and the particle under only exchange in noncontinuous vertical direction.
Therefore the particle has high probability to flow in noncontinuous vertical
line which reflected in the mode dispersion of zigzag stripe a node point. As
the particle momentum increases the exchange flow direction through the
stripe and its edge is reversed and begin to increase again as a reflection
for $\cos^2(\sqrt{3}q_x a/2)$ dependance of the modes dispersions of zigzag
stripe, as $q_x$ reach $\pi$  the modes dispersions reach the maximum
(minimum) energy.

The situation is complectly different for a particle in any edge site in the
zigzag stripe because the edge site has coordination number equal to either
two or one and consequently the particle in the edge site will be under only
one exchange (hopping) force. If the edge site has coordination number equal
to two, the particle in the edge site will be under only the exchange
(hopping) strength in the zigzag line direction and the particle has high
probability to flow in the edge zigzag line, while the exchange (hopping)
strength dependance on the particle momentum $q_x$ is effecting the particle
flow in the zigzag line in this case since no competition with missing
vertical exchange (hopping). Only when the particle momentum reaches near the
value that verify the condition $q_xa/\pi=0.5$ the exchange (hopping)
strength in the zigzag line direction is nearly zero, and the particle become
localized in the edge sites, which create the edge localized states. The
flatness of edge states coming from the small range of $q_x$ around
$q_xa/\pi=0.5$ where the exchange (hopping) strength in the zigzag line
direction at edge sites is nearly zero. Since any small energy delivered to
or taken from the localized particles at edge will move them either to
conduction or valence band the position of localized edge states is the Fermi
Level.

If the edge site has coordination number equal to one, the particle in it
will be under only the exchange (hopping) strength in vertical direction and
therefore the particle will has small probability to flow inside the zigzag
stripe while the particle will have high probability to become localized in
edge sites regardless its momentum $q_x$  which then create an extended flat
edge localized states at Fermi level.

Now we can use the elements of exchange matrix $T(q_x)$ and the Table
\ref{tabexchn} to understand the exchange (the hopping) flow of particles in
armchair stripes. The diagonal elements are equal to $[({SJ}/{2})\exp(-iq_x
a)]$ while up and under elements are equal to $[({SJ}/{2})\exp(-iq_x a/2)]$
which reflect that every site in one line of armchair stripe has only one
nearest neighbor site in the same line, up line, and under line as seen in
Figure \ref{sites}, the half of up and under elements is due to the angle
between up and under sites and the vertical of armchair lattice. The complex
nature of armchair exchange matrix $T(q_x)$ elements show that the particle
is forced to rotate from any armchair line to up or down lines due to the
discontinuity in that lines. The particle in any interior site in the
armchair stripe will be under three competitive exchange (hopping) force with
different strengths: one strong through an armchair line along the
translation symmetry of the stripe and the other two with equal less strength
through up or down lines. Due to absence of armchair line contusions, the
particle flow pattern through armchair stripe will have interface effect
\cite{PhysRevB.59.8271} which lead to highest probability to hopping in
aromatic cyclic chains with small interchain hopping probability
\cite{PhysRevB.54.17954,PhysRevB.75.165414,PhysRevB.73.045432,JPSJ.65.1920,PhysRevB.49.8901,PhysRevB.46.3159,H.Hosoya1990},
and the number of those available complete aromatic cyclic chains depends on
the the number of lines in the armchair. At the value of $q_xa/\pi$ between
$0.25$ and $0.5$ the three exchange strength real part reach minimum and the
imaginary part value reach maximum which mean that the particle will be
nearly trapped inside an aromatic cycle, in this case high energy will be
needed to move it to another aromatic cycle in the armchair stripe, which
displayed as large band gap at the three armchair stripes.

While at $q_xa/\pi=0.0$ the three exchange strength are nearly equal to pure
real value which mean that the particle will be propagate inside an armchair
line parallel to the direction of translation symmetry of armchair, to move
the particle from armchair line in an aromatic cyclic chains to anther chain
an energy will be needed which depend on the aromatic cyclic chains pattern
guided by the texture of the ring currents under applying week magnetic field
perpendicular to graphene nanoribbons shown in Figure \ref{current2} given in
reference \cite{PhysRevB.59.8271}. The Figure show that the armchair has
three aromatic cyclic chains patterns for the three armchair types 3i, 3i+1,
and 3i+2. It is clear that the particle at armchair type 3i+2 has great
probability to tunnel from one chain to anther chain, since they are
connected especially near the edge of the stripe which shown as touching
between the conduction band and valence band at the Dirac point in the stripe
dispersion relation. While the probability of tunneling of particle for the
other two armchair types is neglected, and the particle need some energy to
move from one chain to another chain which shown as two different band gaps
between the conduction band and valence band the two stripes dispersion
relations.

In armchair stripe there is only one kind of edge, where sites from
sublattice A is connected with sites from sublattice B, those edge sites has
only two coordination number, and the particle at those edge sites will be
under only two exchange strength, which in that case are always not balance
and consequently the particle will flow in edge armchair line parallel to the
direction of translation symmetry of the stripe regardless its momentum $q_x$
which explain the absence of flat localized edge states in armchair stripe.

The important difference between the particles exchange flow in zigzag and
armchair stripes is the nature of flow as translation or rotational inside
the stripe. While the exchange flow in zigzag stripes is a translation flow
which shown in real nature of zigzag exchange matrix, the exchange flow in
armchair stripes is a rotational which shown in complex nature of armchair
exchange matrix and clarified in the converting it to real equivalent matrix
\cite{Selim2011,Ahmed2}$$\left[%
\begin{array}{cc}
  \mathbf{Re}(q_x) &-\mathbf{Im}(q_x) \\
  \mathbf{Im}(q_x) & \mathbf{Re}(q_x)\\
\end{array}%
\right] $$
where the real part sub matrix is equivalent to
$\mathbf{\cos(\theta)}$ function, and the imaginary part sub matrix is
equivalent to $\mathbf{\sin(\theta)}$ function and therefore it is no more
than a rotation matrix with argument $q_x$. It is important to note that the
flow in the extended graphene is  similar to armchair stripe since the
particles have a real angular momentum described by its pseudospin
\cite{Mecklenburg2011}.

\section{Applying exchange matrix to 2D square Lattice}

\begin{figure}[h]
\centering
\includegraphics[scale=0.5]{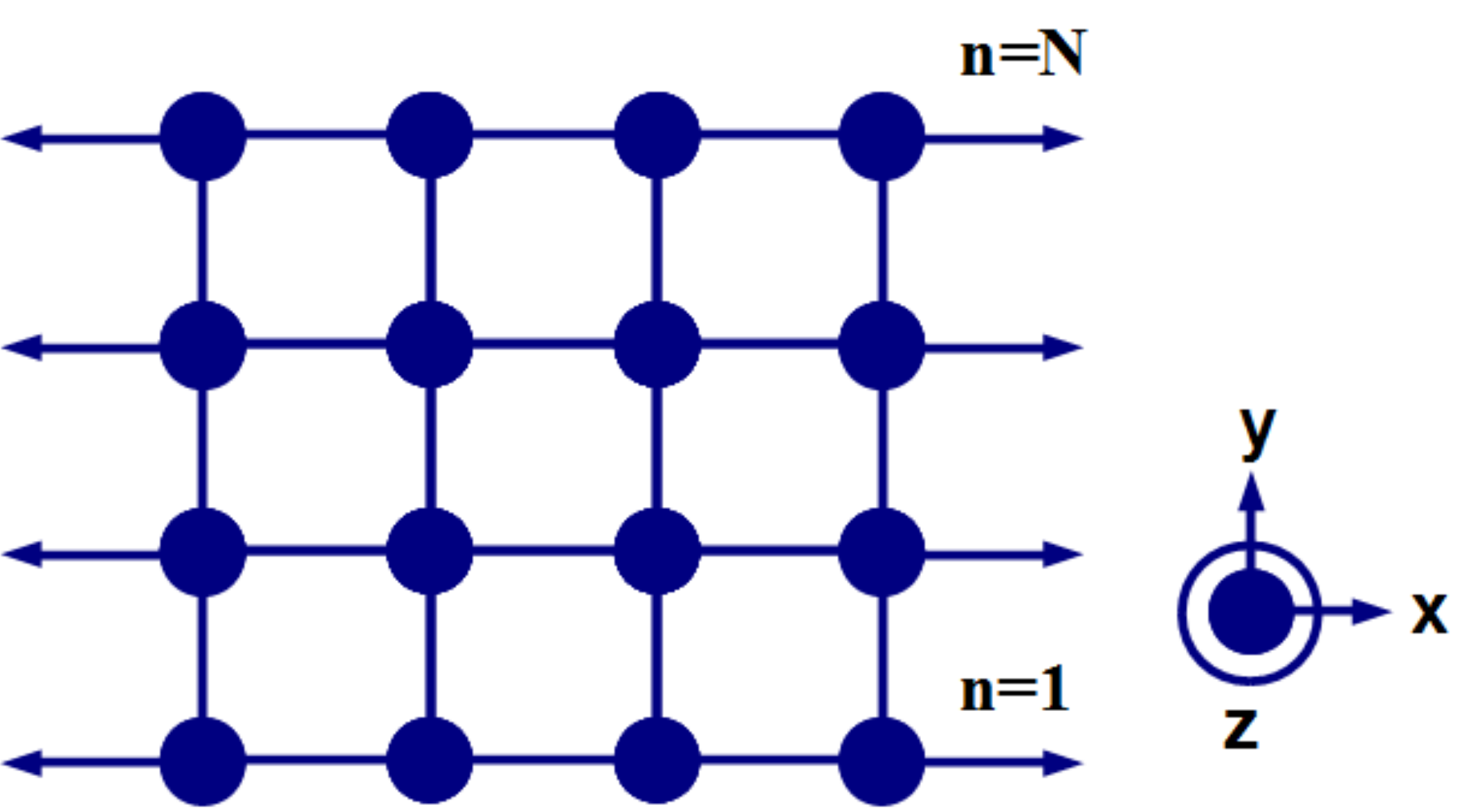}
\caption{Geometry of a 2D Heisenberg ferromagnetic square lattice nanoribbon. The spins are in the $xy$-plane and
the average spin alignment is in $z$ direction. The nanoribbon is finite in $y$ direction with $N$ atomic rows
($n = 1,\cdots,N$).}\label{fig:squarelattice}
\end{figure}
Understanding the exchange matrix can help in the study of the 2D
tight-binding and Heisenberg models for different 2D lattices configurations.
The model easily explains the existence of flat band in 2D lattices and can
be compared to other method \cite{PhysRevLett.106.236803}. We can apply the
exchange matrix to 2D square lattice as following using figure
\ref{fig:squarelattice} to identify the nearest neighbor connections for a
site in 2D square lattice and applying to it the definition of exchange sum
\ref{exsum4}. The obtained exchange matrix for 2D square lattice is given in
Table \ref{tabexchn2}, which is real matrix as expected from the square
lattice geometry. Since the 2D square lattice is Bravais lattice there are
only one lattice sites and therefore the $\mathbf{E}$ matrix size is $N
\times N$ and it is equal to summation of insite energy matrix and the
exchange matrix, i.e. $ \mathbf{E} =\alpha I_N+ T(q_x)$. Actually it is turn
out that $\mathbf{E}$ is the matrix obtained before for
magnetic 2D square lattice using tridiagonal method \cite{Ahmed}.
\begin{table}[h!]
\caption{Nearest neighbor exchange matrix elements for 2D square
lattice}\label{tabexchn2} \centering
\begin{tabular}{ll}
 \hline\hline
    Parameter     &   Square lattice                  \\\hline
   \hspace{50pt}  &  \hspace{50pt}             \\
 $\beta$   &$\frac{SJ}{2}$   \\[10pt]
    $\varepsilon$       &$\frac{SJ}{2} (2\cos(q_x a))$  \\[10pt]
   $\gamma$      &$ \frac{SJ}{2}$  \\
\hline
\end{tabular}
\end{table}

\begin{figure}[hp]
\centering
  \begin{tabular}{cc}
\includegraphics[scale=.6]{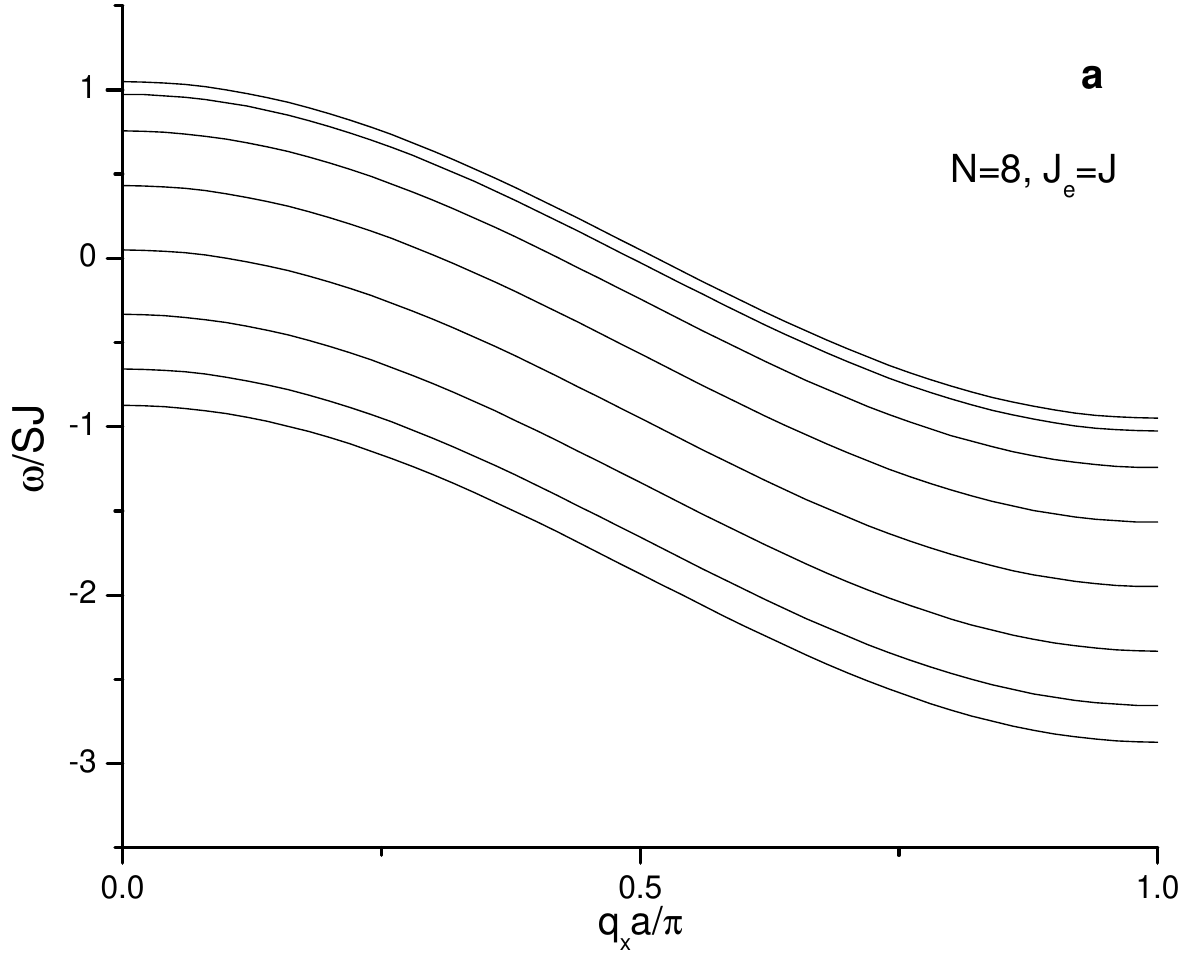}&\includegraphics[scale=.6]{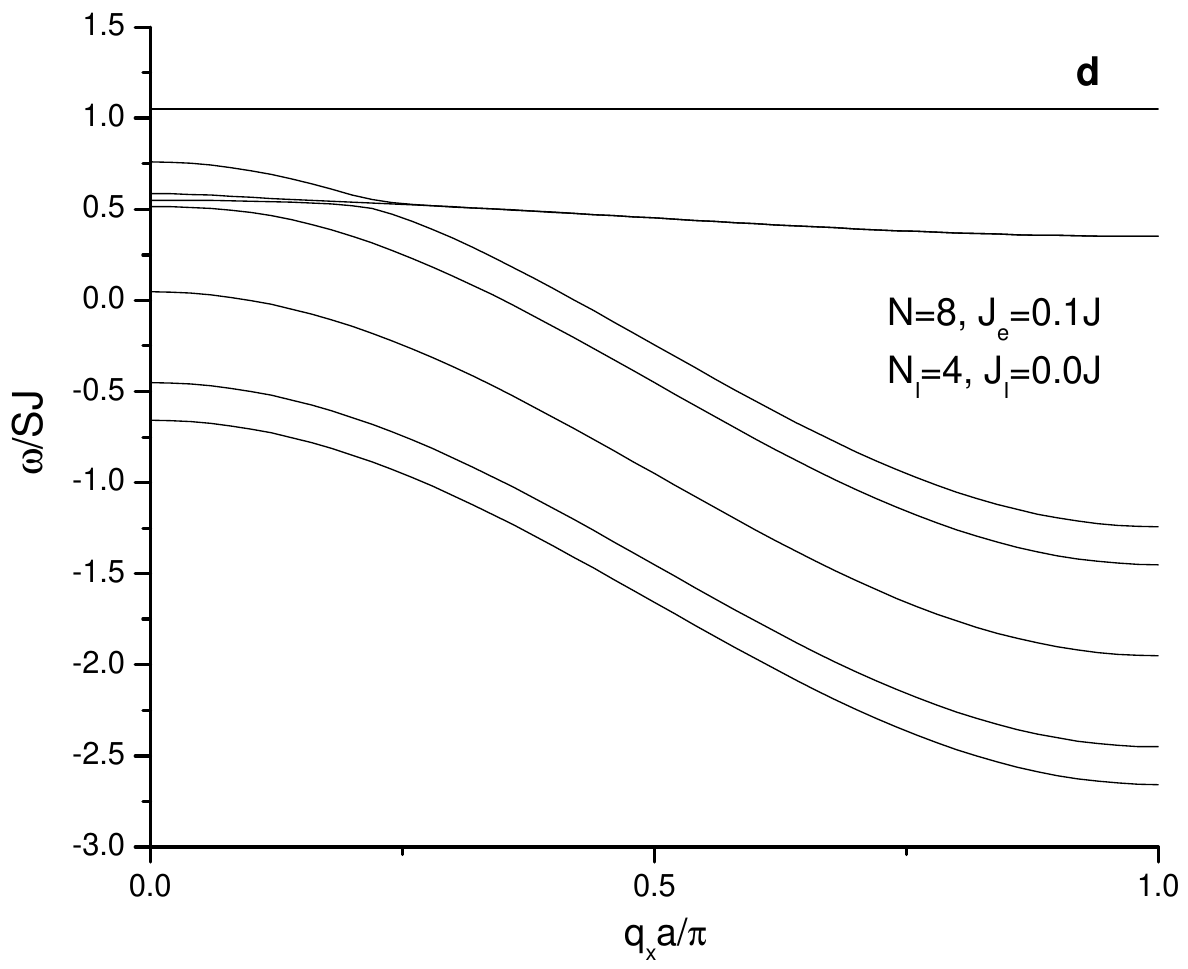}\\
\includegraphics[scale=.6]{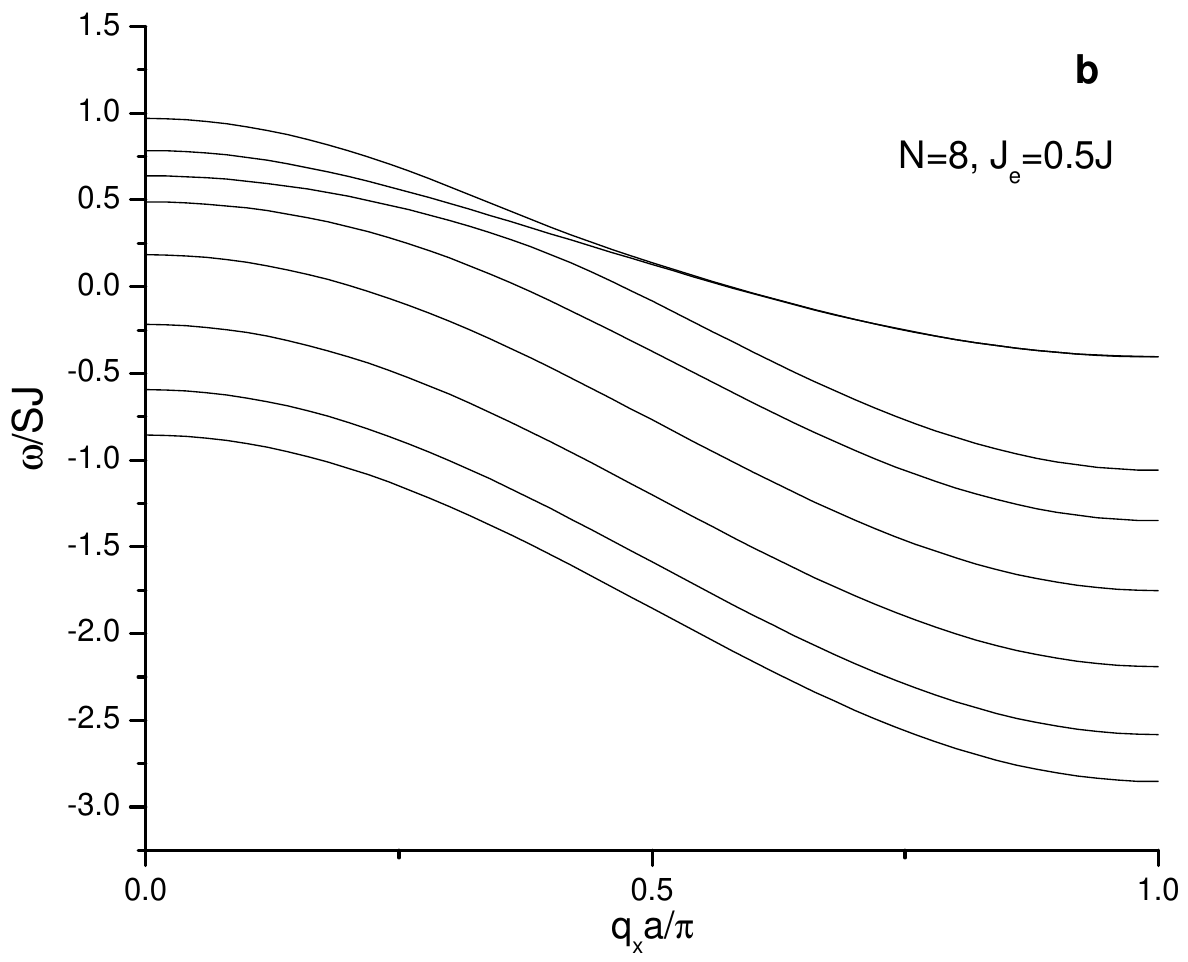}&\includegraphics[scale=.6]{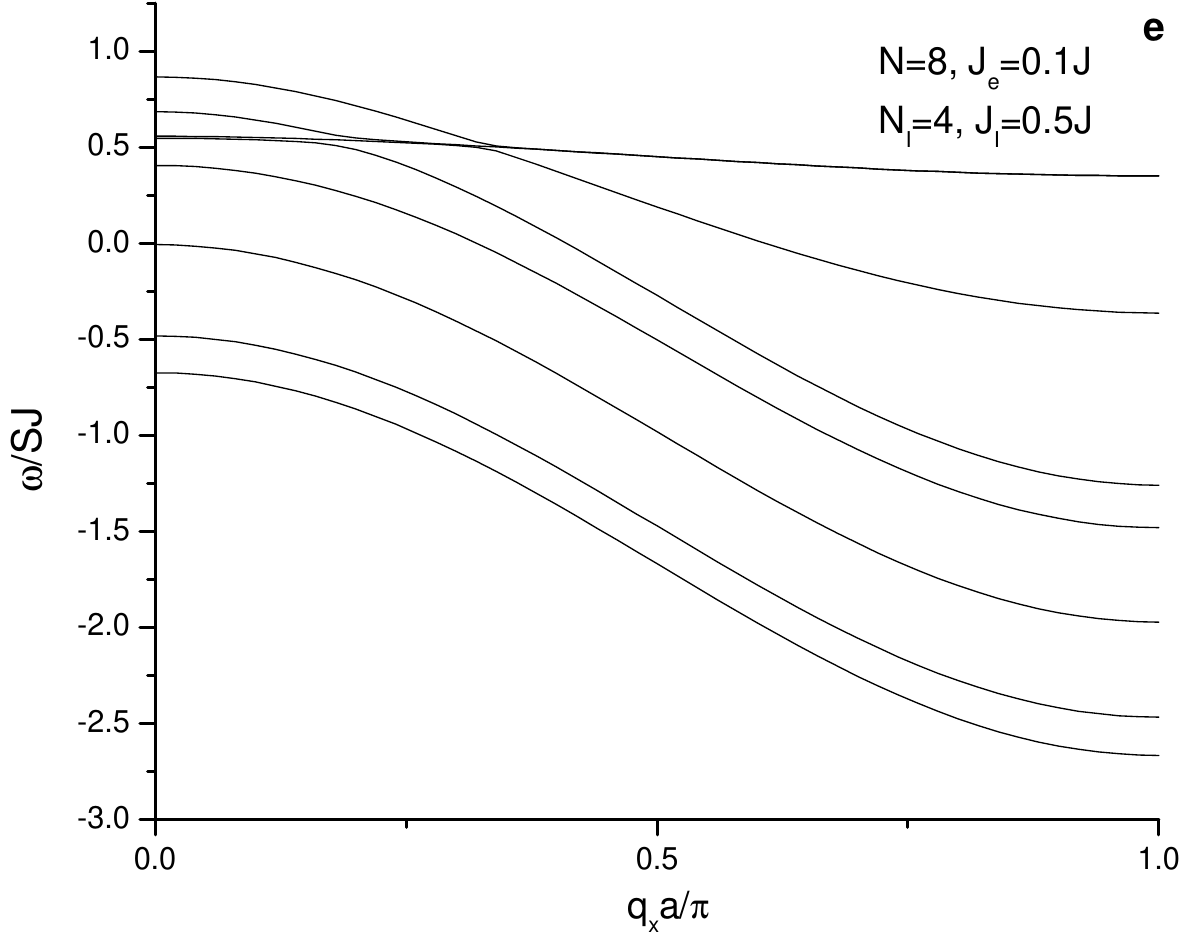}\\
\includegraphics[scale=.6]{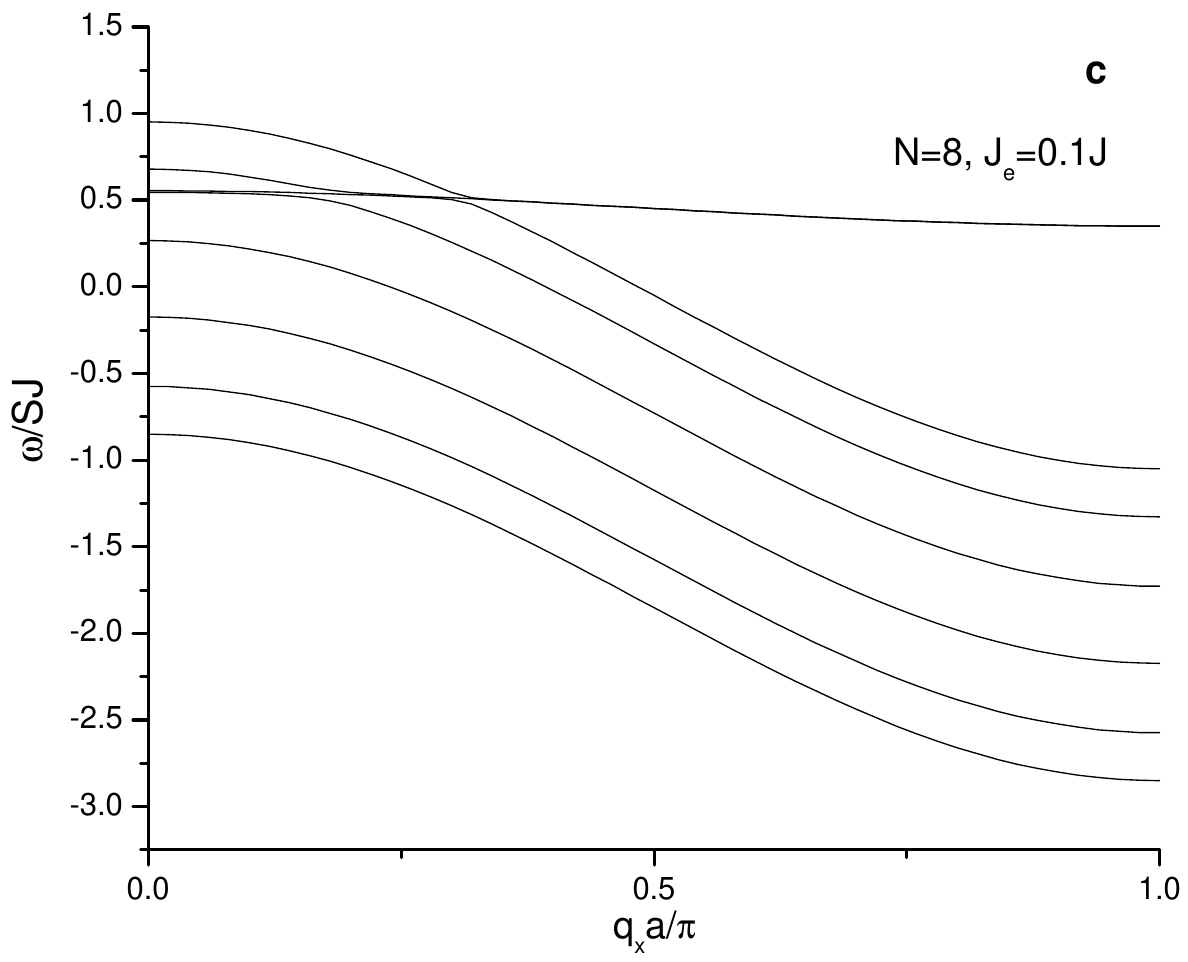}&\includegraphics[scale=.6]{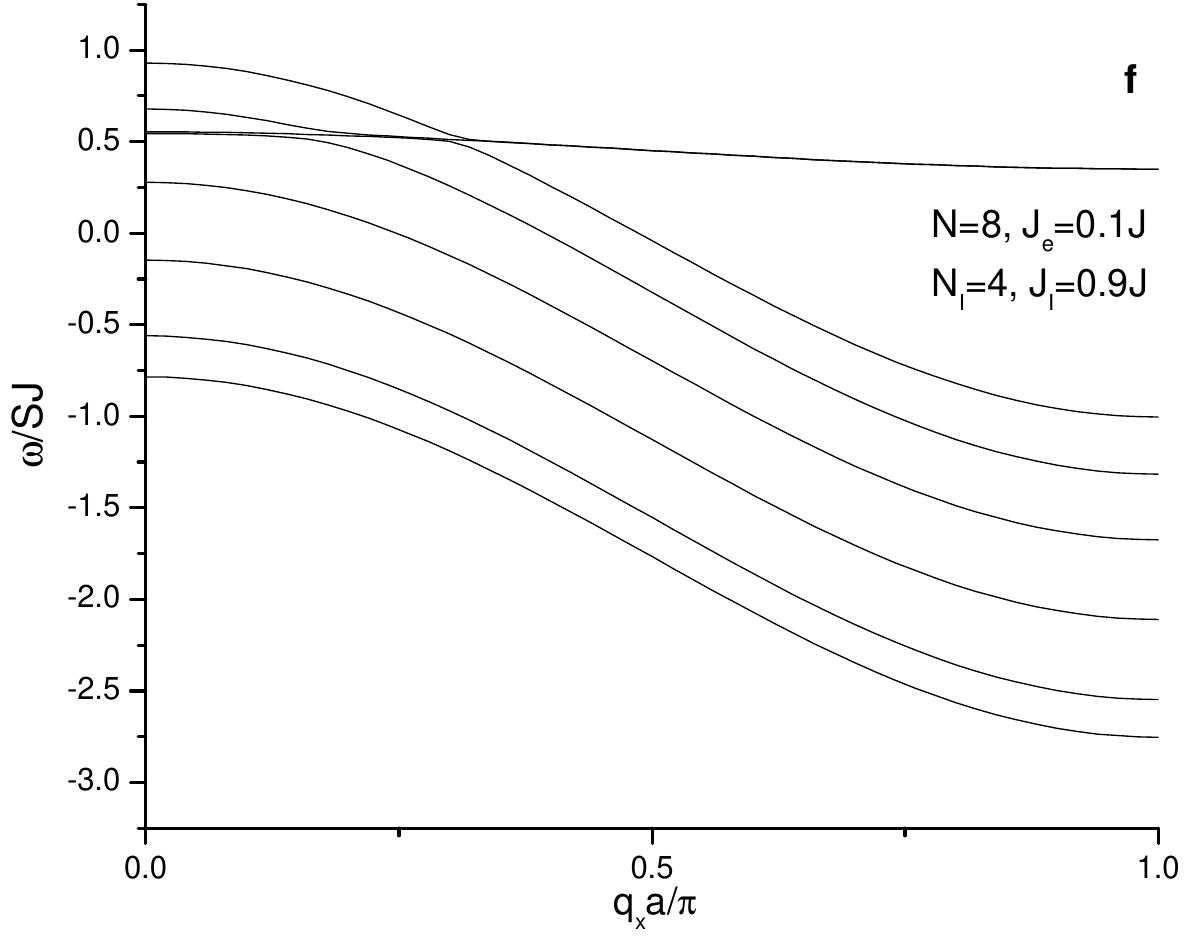}
\end{tabular}
  \caption{Spin waves dispersion for ferromagnetic 2D square
lattice stripe for $N=8$  where
$D=D_e=D_I=1.0$  and $\alpha=-0.95$ (a) $J_e=J$ (b) $J_e=0.5J$ (c) $J_e=0.1J$. Adding an impurity line at line number 4 for $J_e=0.1J$ with (d) $J_I=0.0J$ (e) $J_I=0.5J$ (f) $J_I=0.9J$.}\label{2Dsquare}
\end{figure}

Figure \ref{2Dsquare} shows the obtained spin wave dispersions for
ferromagnetic 2D square lattice stripe with width $N=8$. The right hand side
is describing a magnetic stripe without impurities and with different edge
exchange while the left side describes a magnetic stripe with an impurity
line at line number 4 and with different impurity exchange.  The figures show
the unexpected feature of ferromagnetic 2D square lattice  that the area and
edge spin waves only exist as optic mode as seen before in \cite{Ahmed}, which
now can be understand form the exchange matrix for 2D square lattice in Table
\ref{tabexchn2}. The  diagonal element $(SJ/2) (2\cos(q_x a))$ shows that in
2D square lattice, the exchange strength for nearest neighbor between sites
lies in the same line along the direction of translation symmetry of the
stripe is larger compared by continues exchange strength for vertical nearest
neighbor sites for most values of $q_x$.  Since there is only one lattice
sites type, the exchange flow in all lines is parallel and the main rule of
the exchange in vertical nearest neighbor sites is to quantizing and reducing
the energy as $q_x$ decreases in every mode. Unlike the two sublattice zigzag
case, there is no localized edge states at $q_xa/\pi=0.5$ due to the absence
of the two sublattice in the 2D square lattice and consequently its
determinant depending on the exchange matrix $T(q_x)$ which lead to $\cos(q_x
a)$ dependance of its modes dispersions.

Figure \ref{2Dsquare}a shows the dispersion when the two edge exchange are
equal to interior sites exchange, which for given material properties lead to
absence of edge modes for the 2D square lattice, as the two edge exchange
begin to decrease with respect to interior sites exchange, the strength for
nearest neighbor exchange between sites lies along the edge begin to
decreases which have more effect on particles with the low energy.

Figure \ref{2Dsquare}b shows the effect of reducing edge exchange to half the
value of the interior sites exchange. The particles with low energy become
more localized on the edges, and less able to exchange with interior sites
which make the two edges modes become degenerate and become outside the area
modes boundary at low energy. While the particles with high energy still able
to exchange with edge and interior sites, which show as no effect on the
edges modes at high energy.

Figure \ref{2Dsquare}c shows the effect of reducing edge exchange to 0.1 the
value of the interior sites exchange. The particles with most $q_x$  values
become nearly localized on the edges, and almost not able to exchange with
interior sites which make the two edges modes become flat degenerate outside
the area modes boundary. The total energy of the two localized edges modes
redistributed to equalize the particles energy residue on them, which lead to
increase the energy of localized edges mode. The result is a large nearly
flat edge mode, its energy are very near from high energy of the nearest
neighbor upper and lower lines next to the edges and due to the coupling of
the two edges with those two interior lines through vertical exchange, a
resonance acquire between the edges flat mode and the high energy region of
those two interior modes as seen in the figure. While the particles with high
energy in edges modes are still able to exchange with edge and interior
sites, which show as no effect on the edges modes at high energy.

Figure \ref{2Dsquare}d shows the modified dispersion relations due to the
effect of introducing substitutional a magnetic impurities line at row 4. The
introducing of the impurities line have the effect as the case of zigzag
stripes which is splitting the stripe to two interacted substripes with 3
lines and 4 lines. The strength of the interaction between the two sub
stripes depend on the value of the impurities exchange value $J_I$ , the
Figure shows case when $J_I$ = 0, in this case the expanded impurities flat
localized states appear above the the area modes boundary. Those localized
states are understood as accumulation sites for magnons in the interface
created by the tunneling between the two substripes through the impurities
line, and in the 2D square lattice only particles with highest energy will be
able to tunnel through the impurities line, which shown as absence of highest
energy mode form without impurities area modes.

Figure \ref{2Dsquare}e shows the modified dispersion relations due to
increasing of impurities lines exchange from zero to 0.5 from interior
exchange, the Figure show that particles with high energy begin to flow
between the two substripes and their energy mode part enter to the area modes
boundary, while particles with low energy part from the impurity mode become
localized flat branch outside the area modes.

Figure \ref{2Dsquare}f shows the modified dispersion relations due to
increasing of impurities lines exchange from 0.9 from interior exchange, the
Figure show that particles with nearly all value of energy begin to flow
normally between the two substripes and their inter mode enter to the area
modes boundary, i.e. the stripe become nearly without impurity

\section{Discussion and Conclusions}
In this work, a trial for understanding is done to the construction of
exchange (hopping) matrix for short range (nearest neighbor) interaction by
its lattice geometrical effect on particles flow (its topology).

This is used to explain the dispersion relations for 2D honeycomb lattice
with zigzag and armchair edges obtained for graphene nanoribbons and magnetic
stripes. The explanation shows the rule of zigzag edge geometry
\cite{Tao2011} in the appearance of peculiar localized edge states, and
explain its absence in case of armchair edge configuration.

Using this understanding to construct the exchange matrix for 2D square
Lattice and study the edge and impurities effects on its dispersion
relations, the exchange matrix is used to give a physical interpretation for
obtained results. The obtained results for 2D square Lattice using exchange
matrix shows a similar behavior for its results obtained using the
tridiagonal method discussed in \cite{Ahmed}.

Despite the fact that the exchange method gives very reasonable physical
explanation for the 2D square Lattice results, the tridiagonal method has
more advantage in study the edge effects and its energy states due to the
separation of edges modes from area modes as shown in \cite{Ahmed}. This shows
the needs for study the edge states of 2D honeycomb lattice with zigzag edge
using tridiagonal method, as it is done in \cite{Ahmed3}.

\begin{acknowledgments}
This research has been supported by the Egyptian Ministry of Higher Education
and Scientific Research (MZA).
\end{acknowledgments}

\bibliography{xbib2}

\end{document}